\documentclass[a4paper]{article}

\usepackage{INTERSPEECH2022}

\title{Complex-Valued Time-Frequency Self-Attention for Speech Dereverberation}
\name{Vinay Kothapally, John H.L. Hansen}
\address{Center for Robust Speech Systems (CRSS), The University of Texas at Dallas, TX, USA}
\email{vinay.kothapally@utdallas.edu, john.hansen@utdallas.edu}

\begin{document}

\maketitle
\begin{abstract}
  Several speech processing systems have demonstrated considerable performance improvements when deep complex neural networks (DCNN) are coupled with self-attention (SA) networks. However, the majority of DCNN-based studies on speech dereverberation that employ self-attention do not explicitly account for the inter-dependencies between real and imaginary features when computing attention. In this study, we propose a complex-valued T-F attention (TFA) module that models spectral and temporal dependencies by computing two-dimensional attention maps across time and frequency dimensions.  We validate the effectiveness of our proposed complex-valued TFA module with the deep complex convolutional recurrent network (DCCRN) using the REVERB challenge corpus. Experimental findings indicate that integrating our complex-TFA module with DCCRN improves overall speech quality and performance of back-end speech applications, such as automatic speech recognition, compared to earlier approaches for self-attention.
\end{abstract}

\section{Introduction}
Speech dereverberation is a method for eliminating the ambiguous effects introduced by surround reflections when the speech is captured by a distant microphone. Statistical-based speech enhancement techniques played a crucial role as a front-end processor in several speech processing pipelines, such as \cite{dereverb1, dereverb2, dereverb3, GTSAD}. The requirement for such robust speech systems to assist in the recognition of naturalistic speech recorded by distant microphones has become more important as human-machine interaction technologies gain traction \cite{challenge1,challenge2,fs01, fs03}. Additionally, advances in technologies such as hearing aids require the speech systems to enhance perceptual quality of speech captured in adverse environmental conditions, thus improving human hearing abilities. Several deep learning (DL)-based speech enhancement systems have been successfully developed to address concurrent improvements in perceptual quality and performance of back-end speech and language applications using fully convolutional neural networks (FCN), and recurrent networks (RNN) \cite{monauralTCN,monauralDNN,monauralWRN,monauralGCRN}. The majority of these approaches work with the complex short-term fourier transform (STFT) of distorted speech, either to enhance the log-power spectrum (LPS) and reuse the unaltered distorted phase signal \cite{Mag_1, Mag_2, Ernst,skipconvnet,skipconvgan}, or to estimate the complex ratio mask (cRM) \cite{CRM_1,CRM_2,crm3} and directly enhance the complex spectrogram to restore a cleaner time-domain signal. Enhancing the magnitude response or LPS enables back-end speech applications to operate more efficiently. This is because the most of of back-end speech applications are trained using LPS-derived speech features. Alternatively, speech applications aimed at enhancing perceived quality and intelligibility of speech make extensive use of complex spectrograms to recover magnitude \& phase of distorted signal using DL approaches.

As deep neural networks (DNN) advance to be compatible with complex representations, researchers have investigated many speech enhancement strategies to estimate cRM using deep complex neural networks (DCNN). To address reverberation which distorts the signal in both time and frequency, many sequence-to-sequence learning strategies such as recurrent neural networks (RNNs) and long short-term  memory (LSTM) \cite{LSTM, monauralWRN} have also been explored. In addition to the FCNs, these methods capture and leverage the temporal correlations for speech  dereverberation. In recent years, self-attention (SA) has become a widely utilized mechanism for sequence-to-sequence learning tasks \cite{attn1,attn2,selfattn_conv,attn_speech1}. SA is a mechanism for selective context aggregation that generates an output sequence by computing a weighted average of the input sequence. The learned weights represent the level of attention the network pays to subsets of the input sequence while generating an output sequence. For speech dereverberation task, SA will allow the network to attend time-frequency (TF) locations to reduce the smearing effects of reverberation. However, conventional SA approaches used in DCNN networks do not account for the inter-dependencies between real and imaginary component of complex-valued features. 


The purpose of this study is to develop a complex-valued time-frequency (T-F) self-attention mechanism that computes attention using both real and imaginary components to accurately model temporal dependencies using deep neural networks. To demonstrate the effectiveness of our proposed complex-valued SA mechanism, we integrate two SA approaches with DCCRN: (i) the conventional self-attention mechanism \cite{attn1}, and (ii) the sample independent dual attention block (SDAB) \cite{SDAB} using channel-wise concatenated real and imaginary components. The REVERB challenge corpus is used to examine the improvements in overall speech quality and back-end speech application performance achieved by integrating our proposed self-attention with a fully convolutional and recurrent network, DCCRN.


\section{Problem Formulation}
For a given acoustic environment, a speech signal received by an omni-directional microphone can be modeled as:
\vspace{-0.5em}
\begin{equation} 
\begin{aligned}[b]
x(t) = \sum_{t=0}^{L}{s(t)h(L-t)} + n(t)  \in \mathbb{R}\\ 
X(t,f) = H(t,f)S(t,f) + N(t,f)  \in \mathbb{C}
\label{eq:reverb}
\end{aligned}
\end{equation}
\vspace{-0.5em}

\vspace{-0.5em}\noindent where $x(t)$ is the signal as observed by a distant microphone, $s(t)$ is the clean speech signal from the source, $n(t)$ is additive background noise, $h(t)$ is the room impulse response (RIR) from the source to the microphone, and $L$ represents number of samples in RIR. The relation in frequency domain can be represented as Eq-(\ref{eq:reverb}), where $X(t,f)$, $S(t,f)$, $H(t,f)$ and $N(t, f)$ represent the STFT of observed noisy and reverberated speech, clean speech from source, RIR, and background noise respectively. 
The goal of speech dereverberation is to estimate the complex spectrogram, $\hat{S}(t,f)$ from $X(t,f)$. For this, we estimate a cRM using DCNN to jointly estimate real and imaginary components of enhanced speech, as shown in Eq-(\ref{eq:masking}),

\vspace{-0.5em}
\begin{equation} 
\begin{aligned}
\hat{S}(t,f) = \mathbf{\Psi}(X(t,f))X(t,f) = M(t,f)X(t,f), \\ 
\hat{S_r}+j\hat{S_i} = (M_rX_r-M_iX_i) + j(M_rX_i+M_iX_r) \label{eq:masking}
\end{aligned}
\end{equation}
\vspace{-0.5em}

\noindent where, $M(t,f)$ is the CRM estimated by a DCNN ($\mathbf{\Psi}$). $M_r$, $M_i$, $X_r$, $X_i$, $\hat{S_r}$, $\hat{S_i}$ are the real and imaginary components of the estimated CRM, complex spectrograms corresponding to reverberant and enhanced speech respectively.

\section{Complex-Valued Deep Network}
In this section, we describe the DCCRN model used as a base architecture to evaluate the self-attention mechanisms. Unlike a real-valued network, a complex network allows us to associate each intermediate feature map with its real and imaginary components. Complex convolutions within a complex network perform real-valued convolutions on real and imaginary components of the feature maps and weights. 

\begin{figure}[h!]
\vspace{-1em}
\flushleft
  \includegraphics[width=\linewidth]{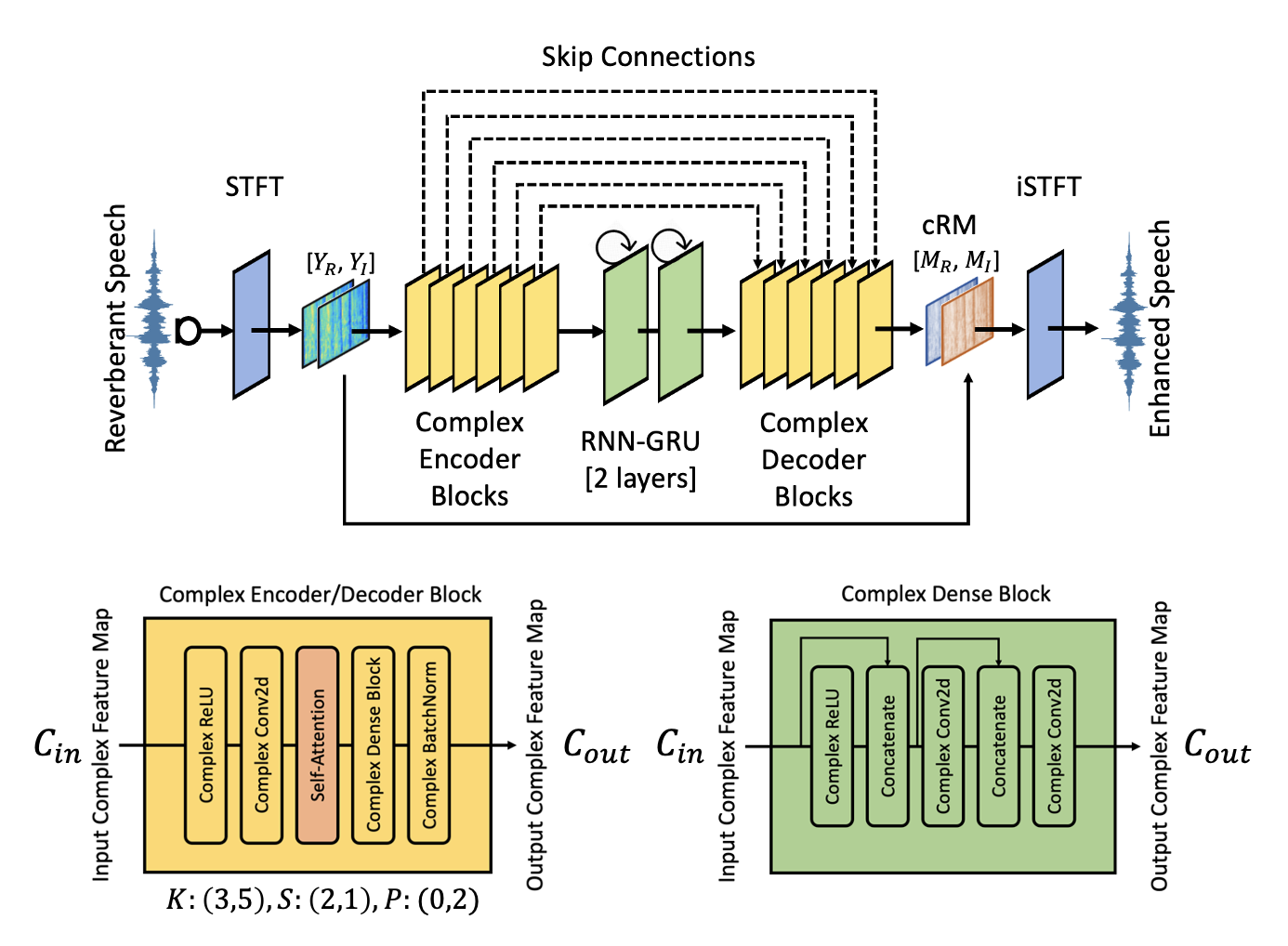}
  \vspace{-1.5em}
  \caption{Deep Complex Convolutional Recurrent Network (DCCRN) used as base architecture to compare various self-attention mechanisms}
  \vspace{-0.5em}
\label{fig:network}
\end{figure}

Consider $X$, a feature map with dimensions $T_{in}\times F_{in}\times C_{in}$ which is provided as input to a complex 2-D convolution layer to produce an output feature map $Y$ with dimensions $T_{out}\times F_{out}\times C_{out}$. In order to perform a complex equivalent of conventional convolution, we assume the kernel $W$ to be a complex-valued weight matrix with its corresponding real and imaginary counterparts represented by two real-valued matrices: $A$ and $B$. The output of the complex convolution $(Y)$ is computed as shown in Eq-(\ref{eq:cmplxconv})
where $X_r$ and $X_i$ are the first $C_{in}/2$ channels and remaining channels representing the real and imaginary components of the input feature maps. 

\vspace{-0.7em}
\begin{equation} 
\begin{aligned}[b]
Y&=W*X = (A+jB)*(X_r+jX_i) \\ 
&= (A*X_r-B*X_i) + j(A*X_i+B*X_r)
\end{aligned}
\label{eq:cmplxconv}
\end{equation}

Similarly, a complex-valued versions of activation functions ($\mathbb{C}$ReLU) and normalization ($\mathbb{C}$BatchNorm) techniques are used to design our DCCRN model. For further details on complex modules, we recommend readers to refer \cite{cmplxpaper, pandey}. A detailed block diagram of our DCCRN model used in this study is shown in Fig. \ref{fig:network}. Here, `$C_{in}/C_{out}$' represent input/output channels of an encoder/decoder block, $K$, $S$,and $P$' represents the kernel size, stride and padding parameters used for convolution layers. We use a 6-layer U-Net \cite{UNet}, which is an encoder-decoder network with two layers of gated recurrent units (GRU). The encoder extracts spectral and temporal features from an input complex spectrogram, and the decoder constructs an enhanced complex spectrogram from the encoded features. The real-valued convolutions within each encoder and decoder layer of a conventional U-Net are replaced with their complex counterparts. Each encoder and decoder block consists of a complex-valued ReLU activation, complex-valued convolutional layer, a self-attention module , a complex-valued dense block, and a complex-valued normalization, see Fig. \ref{fig:network}. In this study, we replace the self-attention module within the encoder and decoder with alternate existing versions and the proposed method. The remaining pipeline is left unaltered for a fair comparison.

\vspace{-0.5em}
\section{Attention Mechanisms}
In this section, we describe two alternate approaches to self-attention mechanisms and propose a fully complex self-attention mechanism. Similar to \cite{SDAB}, we use attention mechanisms to estimate attention maps for time and frequency in parallel to address the smearing effects of reverberation.

\subsection{Sample-Independent Dual Attention Block (SDAB)}
\begin{figure}[h!]
  \centering
  \includegraphics[width=\linewidth]{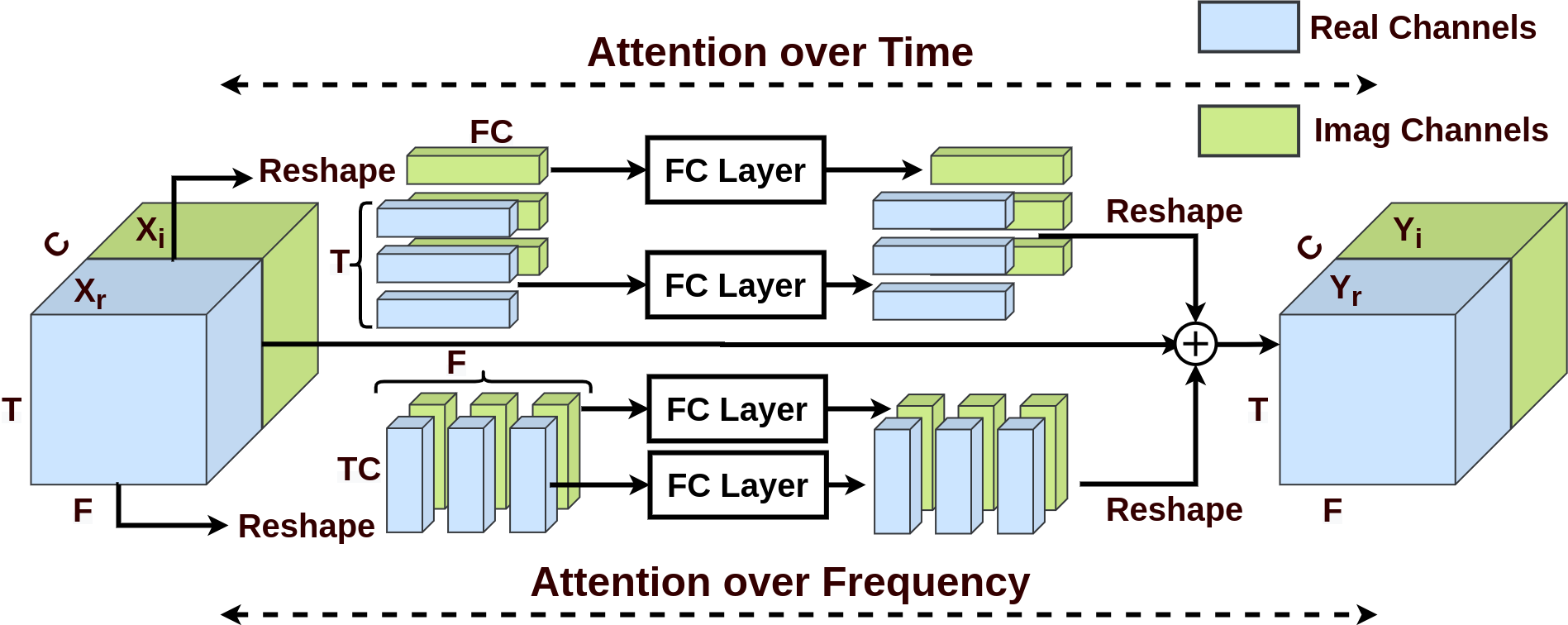}
  \caption{Sample-Independent Dual Attention (SDAB) computed independently for real and imaginary components over time and frequency.}
  \label{fig:sadb}
\vspace{-1em}
\end{figure}

From early speech enhancement studies, lost harmonics along the frequency axis were regenerated using uniform non-linear functions. Likewise, non-linear recursive relations along the time axis are used to estimate signal-to-noise (SNR). Based on this, in \cite{SDAB}, a sample-independent dual attention block (SDAB) was recently proposed, see Fig. \ref{fig:sadb}. Unlike conventional self-attention mechanisms, SDAB estimates attention using fully-connected (FC) layers. Intermediate feature maps are reshaped into stacks of 1-D vectors along the time and frequency axis to form $T\times FC$ and $F\times TC$ matrices. Later, fully-connected layers are used to learn weights for each vector based on their correlations with others along a given dimension. These weights are analogous to the weights learned by a conventional self-attention mechanism with the exception that the sum of weights might not add up to one.  In \cite{SDAB}, real-valued convolutions are employed on complex feature maps stacked as channels. However, in this study we use complex convolutions and compute the SDAB attention mechanism on real and imaginary parts of a complex feature map in parallel.
\vspace{-0.75em}

\begin{figure*}[htb!]
  \centering
  \includegraphics[width=0.85\textwidth]{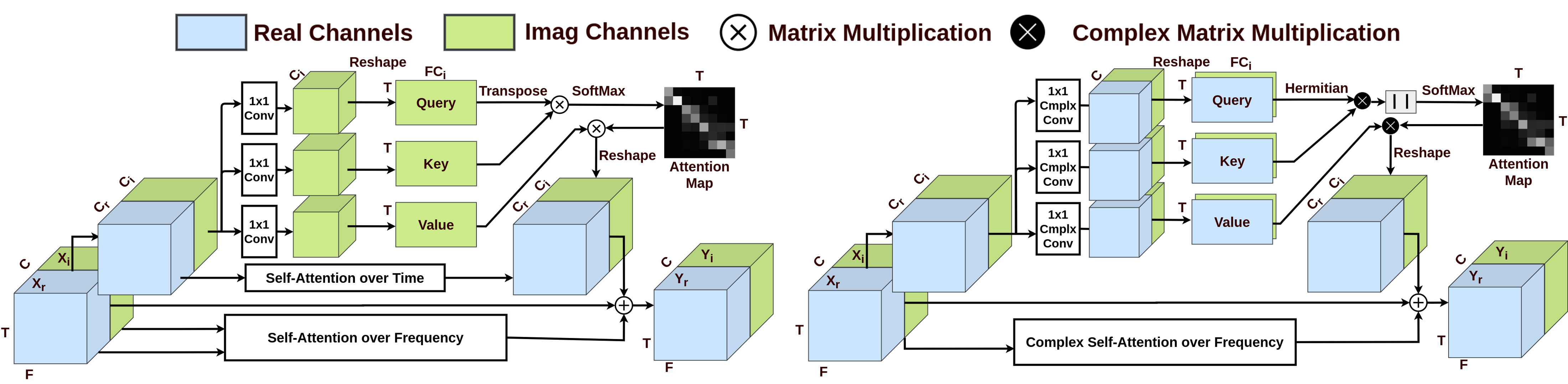}
  \vspace{-0.5em}
  \centerline{(a)  \;\;\;\;\;\;\;\;\;\;\;\;\;\;\;\;\;\;\;\;\;\;\;\;\;\;\;\;\;\;\;\;\;\;\;\;\;\;\;\;\;\;\;\;\;\;\;\;\;\;\;\;\;\;\;\;\;\;\;\;\;\;\;\;\;\;\;\;\;\;\;\;\;\;\;\;\;\;\;\;\;\;\;\;\;\;\;\;\;\;\;\;\;\;\;\;\;\;\;\;\;\;\;\;\;\;\;\; (b)} 
  \caption{Attention computed over time and frequency for complex domain: (a) conventional SA computed independently for real and imaginary components, (b) proposed complex-valued SA computed using real and imaginary components}
  \label{fig:selfattention}
  \vspace{-1em}
\end{figure*}

\subsection{Conventional Self-Attention (SA)}
\vspace{-0.5em}
A conventional self-attention method used for speech applications would ideally map each T-F bin by estimating the contributions of every time-frequency (T-F) bin in a spectrogram. However, in this study, similar to the previous strategy (SDAB), we use a conventional self-attention mechanism to learn the contributions of 1-D vectors along time and frequency axes, see Fig. \ref{fig:selfattention}-a. The self-attention mechanism is a three step process performed on its three major components: query (\textit{\textbf{Q}}), key (\textit{\textbf{K}}) and value (\textit{\textbf{V}}) matrices which are linear projections of the input sequence, see-Eq-(\ref{eq:attention}). First, correlations between query and key are computed using an outer product. Later, these correlations are converted into contributions using a  ``\textit{SoftMax}'' function which results in an attention map. Each row of this attention map represents the contribution of all rows in keys towards a particular row of the query matrix. Finally, this attention map is used to linearly combine rows of value matrix to obtain each row of the output.

\vspace{-1em}
\begin{equation}
\vspace{-0.5em}
Corr = QK^T, \quad
W = \frac{\exp^{Corr(i,j)}}{\sum_{j}^{}{\exp^{Corr(i,j)}}},  \quad
A = WV
\label{eq:attention}
\end{equation}
\vspace{-0.5em}

We use $1\times1$ real-valued convolutions for the linear projections of intermediate feature maps. The real and imaginary parts of these linear projections are then reshaped to construct $T\times FC_{r/i}$ and $F\times TC_{r/i}$ dimensional query, key, and value matrices which are used to compute attention over a given axis. Here, $C_{r,i}$ represents the number of real and imaginary channels of an intermediate feature map. Fig. \ref{fig:selfattention}-a illustrates the self-attention over the time axis for only the imaginary component. Similar computations are carried out in other branches.

\subsection{Proposed Complex-Valued Time-Frequency SA}
A self-attention mechanism, when computed separately for real and imaginary components, does not completely take into account the inter-dependencies of these components in a complex domain. Capturing these inter-dependencies is particularly important for speech applications that need real and imaginary components to be improved jointly to reverse distortions present in the phase. Thus, modifying these components individually might not be an efficient solution for systems aiming at improving the speech quality. Therefore, we propose a fully complex self-attention mechanism that leverages real and imaginary components to estimate the attention over a given dimension, see Fig. \ref{fig:selfattention}-b. Similar to the conventional self-attention mechanism, linear projections of intermediate features maps are performed using $1\times1$ complex convolutions. These projections are then reshaped to form complex query, key, and values matrices. The complex correlations between query and key in the proposed method are computed using Hermitian and complex multiplication operations, Eq-(\ref{eq:cmplxattn}). Later, a ``\textit{SoftMax}'' operation over the magnitude of the correlation scores is performed to estimate an attention map. Finally, this attention map is applied to the complex value matrix to obtain each row of the output.
\begin{align}
 Corr &= QK^H = (Q_rK_r^T + Q_iK_i^T) + j(Q_iK_r^T-Q_rK_i^T),\nonumber\\
 W &= \frac{\exp^{|Corr(i,j)|}}{\sum_{j}^{}{\exp^{|Corr(i,j)|}}},\quad A = WV_r + jWV_i \label{eq:cmplxattn}
\end{align}

The proposed fully complex self-attention has the same number of trainable parameters as a conventional self-attention mechanism used for real and imaginary individually. We strongly believe that fully complex self-attention mechanism, which accounts for the cross-relation between real and imaginary components, should help to improve performance of a DCCN for speech applications.
\begin{table*}[h!]
  \caption{Improvements in Speech Quality Measures on \textit{EvalData} of REVERB Challenge}
  \vspace{-0.8em}
  \label{tab:Speech_Quality}
  \centering
  \renewcommand{\arraystretch}{1.45}
  \resizebox{\textwidth}{!}{
  \setlength{\tabcolsep}{8pt}
  \begin{tabular}{|l|c|c|c|c|c|c|c|c|c|c|c|c|c|c|c||c|} 
    \hline
    ~ & \multicolumn{15}{c||}{\Large{\textbf{Simulated}}} & \Large{\textbf{Real}} \\[0.5ex] \cline{2-17}
    ~ & \multicolumn{3}{c|}{\textbf{\large{CD ($\downarrow$)}}} & \multicolumn{3}{c|}{\textbf{\large{LLR ($\downarrow$)}}} & \multicolumn{3}{c|}{\textbf{\large{FWSegSNR ($\uparrow$)}}} & \multicolumn{3}{c|}{\textbf{\large{PESQ ($\uparrow$)}}} & \multicolumn{3}{c||}{\textbf{\large{SRMR ($\uparrow$)}}} & \textbf{\large{SRMR ($\uparrow$)}} \\[0.5ex]
    
    \textbf{\large{Room}} & \multicolumn{1}{c}{\large{\#1}} & \multicolumn{1}{c}{\large{\#2}} & \multicolumn{1}{c|}{\large{\#3}} & \multicolumn{1}{c}{\large{\#1}} & \multicolumn{1}{c}{\large{\#2}} & \multicolumn{1}{c|}{\large{\#3}} & \multicolumn{1}{c}{\large{\#1}} & \multicolumn{1}{c}{\large{\#2}} & \multicolumn{1}{c|}{\large{\#3}} & \multicolumn{1}{c}{\large{\#1}} & \multicolumn{1}{c}{\large{\#2}} & \multicolumn{1}{c|}{\large{\#3}} & \multicolumn{1}{c}{\large{\#1}} & \multicolumn{1}{c}{\large{\#2}} & \multicolumn{1}{c||}{\large{\#3}} & ~ \\
    \cline{2-17}
    ~ & \multicolumn{16}{c|}{\large{Far Microphone}} \\[0.5ex]
    \hline
    \large{No Processing} & \large{2.672} & \large{5.207} & \large{4.962} & \large{0.518} & \large{0.701} & \large{0.941} & \large{9.781} & \large{6.854} & \large{6.035} & \large{2.621} & \large{2.028} & \large{1.909} & \large{4.586} & \large{2.973} & \large{2.731} & \large{3.175} \\
    \large{WPE} & \large{2.456} & \large{5.163} & \large{4.900} & \large{0.466} & \large{0.678} & \large{0.918} & \large{10.150} & \large{7.103} & \large{6.211} & \large{2.720} & \large{2.082} & \large{1.951} & \large{4.840} & \large{3.204} & \large{2.885}  & \large{3.431}\\
    \large{CplxUNet} & \large{3.632} & \large{4.113} & \large{3.783} & \large{0.510} & \large{0.667} & \large{0.572} & \large{6.872} & \large{6.287} & \large{7.346} & \large{2.567} & \large{2.328} & \large{2.258} & \large{5.277} & \large{4.956} & \large{4.320} & \large{5.343}\\
    \large{SDAB} & \large{2.316} & \large{3.966} & \large{3.754} & \large{0.318} & \large{0.632} & \large{0.667} & \textbf{\large{10.725}} & \textbf{\large{7.594}} & \textbf{\large{7.780}} & \large{2.900} & \large{2.394} & \large{2.325} & \large{5.117} & \large{4.650} & \large{4.163} & \large{4.629}\\
    \large{Conventional SA} & \large{2.177} & \large{3.584} & \large{3.340} & \large{0.244} & \large{0.517} & \large{0.509} & \large{10.771} & \large{6.423} & \large{7.297} & \large{2.920} & \large{2.409} & \large{2.252} & \large{5.370} & \large{5.206} & \large{4.351} & \large{5.648}\\
    \large{Proposed} & \textbf{\large{2.134}} & \textbf{\large{3.548}} & \textbf{\large{3.287}} & \textbf{\large{0.233}} & \textbf{\large{0.513}} & \textbf{\large{0.496}} & \large{9.362} & \large{6.383} & \large{7.777} & \textbf{\large{2.996}} & \textbf{\large{2.451}} & \textbf{\large{2.328}} & \textbf{\large{6.050}} & \textbf{\large{5.335}} & \textbf{\large{4.429}} & \textbf{\large{5.785}}\\
    \cline{2-17}
    ~ & \multicolumn{16}{c|}{\large{Near Microphone}} \\[0.5ex]
    \cline{2-17}
    \large{No Processing} & \large{1.992} & \large{4.634} & \large{4.384} & \large{0.467} & \large{0.452} & \large{0.742} & \large{10.440} & \large{8.712} & \large{7.418} & \large{3.142} & \large{2.419} & \large{2.303} & \large{4.498} & \large{3.746} & \large{3.572}  & \large{3.192}\\
    \large{WPE} & \large{1.854} & \large{4.577} & \large{4.302} & \large{0.444} & \large{0.423} & \large{0.708} & \large{10.694} & \large{8.967} & \large{7.623} & \large{3.289} & \large{2.483} & \large{2.356} & \large{4.626} & \large{3.991} & \large{3.855}  & \large{3.507}\\
    \large{CplxUNet} & \large{3.305} & \large{3.575} & \large{3.562} & \large{0.427} & \large{0.474} & \large{0.465} & \large{8.890} & \large{8.403} & \large{7.835} & \large{2.825} & \large{2.676} & \large{2.591} & \large{5.084} & \large{5.040} & \large{4.831} & \large{5.494}\\
    \large{SDAB} & \large{1.793} & \large{3.575} & \large{3.326} & \large{0.230} & \large{0.445} & \large{0.507} & \textbf{\large{13.359}} & \textbf{\large{10.936}} & \textbf{\large{9.614}} & \large{3.377} & \large{2.750} & \large{2.701} & \large{4.955} & \large{4.828} & \large{4.732} & \large{4.794}\\
    \large{Conventional SA} & \large{1.933} & \large{2.878} & \large{2.817} & \large{0.201} & \large{0.345} & \large{0.386} & \large{10.518} & \large{9.333} & \large{8.899} & \large{3.418} & \large{2.942} & \large{2.702} & \large{5.385} & \large{5.387} & \large{4.662} & \large{5.838}\\
    \large{Proposed} & \textbf{\large{1.904}} & \textbf{\large{2.783}} & \textbf{\large{2.762}} & \textbf{\large{0.200}} & \textbf{\large{0.329}} & \textbf{\large{0.380}} & \large{10.113} & \large{9.561} & \large{9.387} & \textbf{\large{3.479}} & \textbf{\large{3.003}} & \textbf{\large{2.719}} & \textbf{\large{5.506}} & \textbf{\large{5.624}} & \textbf{\large{4.738}} & \textbf{\large{6.012}}\\
    \hline
  \end{tabular} }
  \vspace{-1em}
\end{table*}

\vspace{-0.7em}
\section{Experiments}
All networks studied and compared in this work are evaluated on the REVERB Challenge corpus  \cite{challenge1,challenge2}. The REVERB Challenge corpus is a collection of simulated and real recordings of speech sampled at 16kHz. The simulated data is generated using clean speech from WSJCAM0 and room impulse responses (RIRs) collected from three different sized rooms (small, medium, and large) and two different microphone placements (near, far) for a single microphone, 2-channel, and 8-channel microphone arrays. For further details on the corpus, see \cite{challenge1,challenge2}. Models in this study are trained on 7,861 simulated reverberant and clean utterance pairs which correspond to approximately 15 hours. We use the evaluation set (~5 hrs) of the corpus to compare performances of DCCRN with various attention mechanisms and a widely used statistical dereverberation algorithm,  weighted prediction error (WPE). We use the metrics provided for the REVERB Challenge to evaluate the improvements in speech quality. We recommend referring \cite{challenge1,challenge2,KaldiASR} for better understanding of quality metrics. We also evaluate the improvements in performance for back-end systems such as automatic speech recognition (ASR) and speaker verification (SV) systems by monitoring word error rates (WER) and equal error rates (EER).

\vspace{-0.5em}
\subsection{Experimental Setup}
Figure-(\ref{fig:network}) summarizes the network setup and details of each building block in our system development. The complex kernels are initialized with unitary matrices for better generalization. For a given speech utterance, we generate complex spectral images by first computing STFT with a frame length of 32 ms and 75\% overlap. Next, the lower half of the complex STFT is divided into batches with consecutive frames to form complex spectral images of size $256\times 256$. Similar to our previous work \cite{skipconvnet}, we perform optimal smoothing on power spectral density (PSD) on complex spectral images of reverberant and clean utterance pairs being fed to the network. The output of the DCCRN is a predicted complex mask which is applied to the input (as formulated in Eq-(\ref{eq:masking})) to obtain an enhanced complex spectrogram. In this study, we use a combination of loss functions which are aimed at reducing estimation errors in magnitude and complex domains. We also use dynamic compression on the magnitudes of the clean and estimated signals before computing the loss:

\vspace{-1.2em}
\begin{equation}
\begin{aligned}
  \mathcal{L}_{cmplx} = {} & (1-\beta)\sum_{t,f} \bigl\lvert\lvert S(t,f)\rvert^c -\lvert \hat{S}(t,f)\rvert^c \bigr\rvert^2 \\ 
  & + \beta\sum_{t,f} \bigl\lvert\lvert S(t,f)\rvert^ce^{j\phi_{S}} -\lvert \hat{S}(t,f)\rvert^ce^{j\phi_{\hat{S}}} \bigr\rvert^2, \label{eq:loss}
\end{aligned}
\end{equation}
\vspace{-0.5em}

\noindent where $c$,$\phi_{S}$,$\phi_{\hat{S}}$, and $\beta$ are the compression exponent applied to the magnitudes, phases of clean \& estimated complex spectrograms, and weight factor to combine the magnitude only with the complex loss respectively. Similar to \cite{cmplxloss}, we set both $c$ and $\beta$ to 0.3. All investigated networks in the study are trained using Adam optimizer for 20 epochs with a batch-size of 4. We report improvements seen on performance metrics mentioned in earlier sections for \textit{Eval} set of the corpus.     

In order to evaluate the performance improvements for speaker verification (SV), we train a conventional X-vector extractor \cite{xvectors} on the Voxceleb 1+2 datasets \cite{VoxCeleb1,VoxCeleb2}. Later, a probabilistic linear discriminant analysis (PLDA) model is trained on the clean \textit{Train} set of the REVERB corpus. 
Similarly, to evaluate the performance improvements for an ASR system, we have trained a hybrid DNN-HMM system \cite{KaldiASR} on the clean speech from the \textit{Train} set of the REVERB corpus. Later, we use the original and enhanced speech utterances from the \textit{Eval} set  for evaluations.


\begin{table}[h!]
  \caption{Performance of Back-End Speech Systems on \textit{EvalData}} 
  \vspace{-0.8em}
  \label{tab:Back-end}
  \centering
  \renewcommand{\arraystretch}{1.5}
  \resizebox{\linewidth}{!}{
  \setlength{\tabcolsep}{10pt}
  \begin{tabular}{|l|c|c|c|c|}
    \hline 
    ~ & \multicolumn{2}{c|}{\large{\textbf{Simulated}}} & \multicolumn{2}{c|}{\large{\textbf{Real}}} \\ \cline{2-5}
    ~ & \large{\textbf{WER\%}} & \large{\textbf{EER\%}} & \large{\textbf{WER\%}} & \large{\textbf{EER\%}}\\
    \hline
    \large{REVERB (No Processing)} & \large{38.00} & \large{8.21} & \large{96.04} & \large{6.14}\\
    \large{WPE} & \large{32.65} & \large{8.50} & \large{94.26} & \large{6.14}\\ 
    \large{CplxUNet} & \large{19.15} & \large{7.07} & \large{66.24} & \large{9.94}\\ 
    \large{CplxUNet + SDAB} & \large{13.29} & \large{5.06} & \large{67.93} & \large{6.43}\\ 
    \large{CplxUNet + SA} & \large{11.66} & \large{3.15} & \large{42.51} & \large{5.85}\\ 
    \large{CplxUNet + Proposed FCSA} & \large{\textbf{10.42}} & \large{\textbf{3.05}} &  \large{\textbf{38.42}} & \large{\textbf{5.56}}\\
    \hline
    \end{tabular}
    }
    \vspace{-0.8em}
\end{table}

\vspace{-0.5em}
\subsection{Experimental Results \& Discussion}
We compare the speech quality scores for reverberant speech and complex networks with various self-attention mechanisms in Table-(\ref{tab:Speech_Quality}). Each metric in the table is associated with either a `$\uparrow$' or `$\downarrow$' to represent the metric direction for improvement. We see that irrespective of the type of self-attention mechanism, introducing attention over time and frequency improves overall quality of speech. This can also be confirmed from performance improvements achieved for the back-end systems. Although the SDAB mechanism outperformed all other self-attention mechanisms in improving SNR, it was unable to show similar trends in any other quality metrics. A reasonable conclusion can be made that SDAB improves system performance for speech distorted by additive noise. On a similar note, the self-attention mechanism used in parallel for real and imaginary components could not provide improvements similar to the proposed fully complex self-attention. The proposed attention achieves a $+$\{1.93, 7.59, 18.09\}\% and $+$\{4.28, 27.13, 10.54\}\% relative improvements averaged over all speech quality metrics for simulated and real speech recordings compared to DCCRN with other attention mechanisms. Likewise, Table-(\ref{tab:Back-end}) shows performance of back-end speech systems on speech signals enhanced by various discussed strategies. We see $+$\{10.63, 21.59, 45.5\}\% relative or $+$\{4.09, 29.51, 27.82\}\% absolute improvements on WER for simulated and real speech compared to other networks. Although the corpus was designed to evaluate systems for speech quality and ASR, we used the same \textit{Eval} set from the corpus with limited speaker variability to evaluate speaker recognition system for a fair comparison. Similar to ASR performance, we see that the proposed attention mechanism outperforms others with $+$\{33.24, 20.84\}\% relative improvements in EER.

\vspace{-0.5em}
\section{Conclusion}
In this study, we proposed a fully complex self-attention mechanism for a DCCN which improved the learning ability of the DCCN to map complex reverberant spectrograms to their anechoic counterparts. We compared the proposed method's performance to two distinct approaches to self-attention employed in speech systems. In comparison to alternative SA techniques and a widely utilized WPE dereverberation algorithm, a DCCN coupled with our proposed SA improved speech quality for both real and simulated speech. Additionally, we demonstrated how the proposed attention mechanism benefits back-end speech systems such as ASR and speaker verification. The gains in speech quality and performance of back-end speech applications achieved by the proposed fully complex self-attention mechanism demonstrate that better attention estimations can be computed by our proposed SA which accounts for the inter-dependencies between real and imaginary components of features in the complex domain.

\bibliographystyle{IEEEtran}
\bibliography{refs}


\end{document}